\documentstyle{article}
\font\tenbf=cmbx10
\font\tenrm=cmr10
\font\tenit=cmti10
\font\elevenbf=cmbx10 scaled\magstep 1
\font\elevenrm=cmr10 scaled\magstep 1
\font\elevenit=cmti10 scaled\magstep 1

\renewenvironment{thebibliography}[1]
 { \elevenrm
   \begin{list}{\arabic{enumi}.}
    {\usecounter{enumi} \setlength{\parsep}{0pt}
     \setlength{\itemsep}{3pt} \settowidth{\labelwidth}{#1.}
     \sloppy
    }}{\end{list}}

\begin{document}
\setcounter{page}{0}
\begin{flushright} PSU/TH/139\\
hep-th/9312242\\
December 1993  \end{flushright}

\begin{center}{{\tenbf INSTANTON-INDUCED PARTICLE PRODUCTION \\
               \vglue 10pt
               IN DEEP INELASTIC SCATTERING $\footnote { Talk given at
the XXIII International Symposium on Multiparticle Dynamics}$
\\}
\vglue 1.0cm
{\tenrm IAN BALITSKY $\footnote { On leave of absence from
St.Petersburg Nuclear
Physics Institute, 188350 Gatchina, Russia}$  \\}
\baselineskip=13pt
{\tenit Physics Department, Penn State Univ., 104 Davey Lab.\\}
\baselineskip=12pt
{\tenit University Park, PA 16802, USA\\}
\vglue 0.8cm
{\tenrm ABSTRACT}}
\end{center}
\vglue 0.3cm
{\rightskip=3pc
 \leftskip=3pc
 \tenrm\baselineskip=12pt
 \noindent
I discuss
the instanton-induced contributions
to the coefficient functions in front of parton densities. They
correspond to the spherically symmetric particle production at the
level of $10^{-2}-10^{-5}$ of the total cross section of deep
inelastic scattering from the parton.
\vglue 0.6cm}
{\elevenbf\noindent 1. Introduction}
\vglue 0.2cm
\baselineskip=14pt
\elevenrm
The deep inelastic lepton-hadron scattering at large momentum
transfers $Q^2$ and not too small values of the Bjorken
scaling variable  $x = Q^2/2pq$ is studied in much detail and
presents a classical example for the application of perturbative
QCD. The factorization theorems allow one to separate
the $Q^2$ dependence of the structure functions in coefficient
functions $C_i (x,Q^2/\mu^2 ,\alpha_s(\mu^2))$ in front of parton
(quark and gluon) distributions of leading twist
$P_i (x,\mu^2, \alpha_s (\mu^2))$
\begin{equation}
   F_2(x,Q^2) = \Sigma_i C_i (x,Q^2/\mu^2 ,\alpha_s(\mu^2))
              \otimes
               P_i (x,\mu^2, \alpha_s (\mu^2)),
\end{equation}
where the summation goes over all species of partons, and $\mu$ is the
scale separating "hard" and "soft" contributions to the cross
section. At $\mu^2 = Q^2$ the coefficient functions can be
calculated perturbatively and are expanded in power series in
the strong coupling
 \begin{equation}
 C(x,1,\alpha_s(Q^2)) = C_0(x)+\frac{\alpha_s(Q^2)}{\pi} C_1(x)
     +\left(\frac{\alpha_s(Q^2)}{\pi}\right)^2 C_2(x)+\ldots
 \label{cpert}
 \end{equation}
 whereas their evolution with $\mu^2$ is given by famous
 Dokshitzer-Gribov-Lipatov-Altarelli-Parisi equations.
Going over to a low normalization point $\mu^2 \sim 1 GeV$,
one obtains the structure functions expressed in terms of the
parton distributions in the nucleon at this reference scale.
The parton distributions
absorb all the information about the dynamics of large distances and
are fundamental quantities extracted from the experiment.
Provided the parton distributions are known, all the dependence
of the structure functions on the momentum transfer is calculable
and is contained in the coefficient functions $C_i$.
Corrections to this simple picture come within perturbation
theory from the parton distributions of higher twists and are
suppressed by powers of the large momentum $Q^2$.

The picture described above presents a part of the common wisdom
about hard processes in the QCD, and in a more
or less detailed presentation can be found in any textbook.
Less widely known is the fact that from the theoretical point
of view this picture is not complete. An indication that some
contributions may be missing comes
 from the asymptotic nature of the perturbative series
in (\ref{cpert}). This series is non-Borel-summable, which means
that any attempt to attribute  a quantitative meaning to the
{\em sum} of the series in (\ref{cpert}) would produce an
exponentially small imaginary part
 $\sim i\exp\{-\mbox{\rm const}\cdot \pi/\alpha_s(Q^2)\}$,
 which is to be cancelled by the imaginary part coming from
nonperturbative contributions. Thus, separation between perturbative
and nonperturbative pieces in the cross section as the ones
which contribute to the coefficient function and the parton
distribution, respectively, cannot be rigorous.
A modern discussion of the asymptotical properties of the
perturbation series in QCD can be found in \cite{MUaachen,zakh92}.

In addition to {\em imaginary} exponential corrections which must
cancel identically against the corresponding nonperturbative
contributions, the coefficient functions may acquire also {\em real}
exponential corrections, which potentially produce observable
effects. In this talk I shall report on recent results by
V.Braun and myself \cite{bal93a,bal93b}, indicating that these
corrections are indeed present.
 We have found that
the deep inelastic cross section indeed possesses
exponential contributions of the form
$F(x)\exp[-4\pi S(x)/\alpha_s(Q^2)]$, where $S(x)$, $F(x)$
are  certain
functions of Bjorken $x$, which we are able to calculate in a
certain kinematical domain.
 Since the experimental data
are becoming more and more precise, it is of acute interest to find a
boundary for a possible accuracy of the perturbative approach, which
is set
by nonperturbative effects. Our study has been fuelled by
recent findings of an enhancement of instanton-induced effects
at high energies
in a related problem of the violation of baryon number
in the electroweak theory \cite{ring90,matt92}.
In the case of QCD the instanton-induced effects could
turn out to be significant at high energies,
despite the fact that they correspond formally to  contributions
of a very high fractional twist
$\exp(-4\pi S(x)/\alpha_s(Q^2)) \sim (\Lambda_{QCD}^2/Q^2)^{bS(x)}$.

Note that, in principle,the instanton contributions in QCD are
infrared-unstable -  in a typical situation
integrations over the instanton
size are strongly IR-divergent. However,this problem does not
 affect calculation of instanton contributions
to the coefficient functions which are IR-protected, as we shall see
below (the detailed discussion can be found in \cite{bal93b}).
\vglue 0.4cm
{\elevenbf\noindent 2. Instanton contribution to the structure
function of a gluon}
\vglue 0.4cm
Let us start from the instanton-induced contribution to the
coefficient function in front of gluon density.
Following Zakharov \cite{zakh90}, we calculate the cross section of
the $\gamma^{\ast}g$ scattering
by means of the optical theorem.
 The trick is to evaluate the
contribution to the functional integral
coming from the vicinity of the instanton-antiinstanton
configuration in Euclidian space, and calculate the
cross section by the analytical continuation to Minkowski
space and by taking the imaginary part.
 The true small parameter in our calculation is the value of the
coupling constant at the scale $Q^2$, which ensures that the effective
instanton size is sufficiently small.
We prefer to start, however, from the well-studied situation where not
only the instanton sizes are small, but also the $I\bar{I}$ separation
is much larger than these sizes. As we shall see below, this
corresponds to $1-x\ll 1$ (but
of course $1-x \gg \alpha_s$ still). After that we shall move
to smaller x (which correspond to strongly interacting  $I$ and $\bar{I}$)
 trying to be accurate to collect to the semiclassical
accuracy all the dependence on $\rho^2/R^2$ in the exponent.To this
end,we shall have in mind  the valley method \cite{bal86}, in which
all the dependence on the $\bar I I$
separation is absorbed in the action $S(\xi)$ on the $\bar I I$
configuration. However, we do not take into
account corrections of order $\rho^2/R^2$ in the preexponent,
and to this accuracy
need the first nontrivial term only in the
cluster expansion of the quark propagator at the $\bar I I$
background \cite{andrei}:
 $
  \langle x|\nabla_{I\bar I}^{-2}\bar \nabla_{I\bar I}|0\rangle
 =
 \int dz\, \langle x|\nabla_1^{-2} \bar\nabla_1|z\rangle
 \sigma_\xi \frac{\partial}{\partial z_\xi}\langle z|
 \bar\nabla_2 \nabla_2^{-2} |0\rangle
 $.

 The leading contribution to the gluon matrix
element of the T-product of the electromagnetic currents  is given
 by the following expression
\begin{eqnarray}
\lefteqn{
\int dz e^{iqz}
\langle A^a(p),\lambda |T\{ j_\mu(z) j_\nu (0)\}|A^a(p),\lambda\rangle
^{I\bar{I}}         =}
\nonumber\\
 &&\mbox{} = \sum_q e^2_q
 \int dU
\int \frac{d\rho_1}{\rho_1^5} d(\rho_1)
\int \frac{d\rho_2}{\rho_2^5} d(\rho_2)
\int dR \int dT \int dz e^{iqz}
\nonumber\\
&&\mbox{}\times
\frac{1}{8}
\,\,\mbox{lim}_{p^2\rightarrow 0}\, p^4
\epsilon^\lambda_\alpha \epsilon^\lambda_\beta
Tr\left\{A_{\alpha}^{\bar{I}}(p) A_{\beta}^I(-p)\right\}
e^{-\frac{4\pi}{\alpha_s} S_{\bar I I}}
\nonumber\\
 &&\mbox{}\times
 (a^\dagger a)^{n_f-1}
\left\{ a  \bar \phi_0(0)\bar\sigma_\nu
\langle 0|\nabla_2^{-2}\nabla_2\bar\partial \nabla_1 \nabla_1^{-2}
|z\rangle \bar\sigma_\mu \phi_0(z)
\right.
 \nonumber\\
 &&\mbox{} \left.
 +a^\dagger
\bar \kappa_0(z) \sigma_\mu
 \langle z|\nabla_1^{-2} \bar
 \nabla_1\partial \bar\nabla_2 \nabla_2^{-2}
|0\rangle \sigma_\nu \kappa_0(0)
+(\mu \leftrightarrow \nu, z \leftrightarrow 0)  +\ldots
\right\}
\label{formula1}
\end{eqnarray}
which corresponds to the diagram shown in Fig.1a.
\begin{figure}
    \begin{center}
        \begin{picture}(100,120)
        \end{picture}
    \end{center}
    \caption[xxx]{ \small
                   The instanton-induced contribution to the
                    structure function of a gluon (b) and of a quark
                   (c,d). Wavy lines are (nonperturbative) gluons.
                    Solid lines are
                   quark zero modes in the case that they are
                    ending at the instanton (antiinstanton),
                    and quark propagators at the $\bar I I$ background
                    otherwise.
                    }
   \label{pic.a}
\end{figure}
The full expression contains many more terms \cite{bal93b}
which are not shown because we have found that all of them are
of order $O(\alpha_s(Q^2))$ compared to the expression in
Eq.\ref{formula1}.
The subscript '1' refers to the antiinstanton
with the size $\rho_1$ and the position of the center
 $x_{\bar I} = R+T$,
and the subscript '2'  refers to the instanton with the size
$\rho_2$ and the center at $x_I = T$.
 We use   conventional notations $\nabla = \nabla_\mu\sigma_\mu$ and
$\bar\nabla = \nabla_\mu\bar\sigma_\mu$, etc,  where
$\sigma_\mu^{\alpha\dot\alpha} = (-i\sigma, 1)$,
$\bar\sigma_{\mu\dot\alpha\alpha} = (+i\sigma, 1)$, and
$\sigma$ are the standard  Pauli matrices. Also, we write the quark zero
modes in terms of the two-component Weil spinors
 $\psi_0 = \left(
 \begin{array}{c} \kappa_0\\ \phi_0 \end{array}\right)$,
$\psi^\dagger_0 =
\left( \bar \phi_0\,\, \bar\kappa_0 \right)$,
and $a$ and $a^\dagger$ denote the overlap integrals
 $  a = - \int~dx~ (\bar\kappa \partial \kappa),
 a^\dagger  = - \int~dx~(\bar\phi\bar\partial\phi)$.
Here $ S_{ I\bar I}$ is the action of the instanton-antiinstanton
configuration
 and $\xi=\frac{R^2+\rho_1^2+\rho_2^2}{\rho_1\rho_2}$is the conformal
 parameter \cite{yung88}
(the normalization is such as $S(\xi)=1$ for an infinitely
separated instanton and antiinstanton).
Writing the action as a function of $\xi$ ensures that the
interaction  between instantons is small in two different limits:
for a widely separated $I \bar I $ pair, and for a small instanton
put inside a big (anti)instanton, which are related to each other
by the conformal transformation. In the limit of large $\xi$ the
expansion of $S(\xi)$ for the dominating
maximum attractive $I \bar I $ orientation reads \cite{yung88}
\begin{equation}
  S(\xi) =  \left(1-\frac{6}{\xi^2}+O(\ln(\xi)/\xi^4)\right)
\label{action}
\end{equation}
where the $1/\xi^2$ term corresponds to a slightly corrected
 dipole-dipole interaction.  Thus, the action  $S(\xi)$
decreases with the distance between instantons, so that
the instanton and the instanton effectively attract each
other. This attraction results in the exponential increase
of the cross section --- the effect found by Ringwald \cite{ring90}.
Further terms in the expansion of the action can be obtained by the
so-called valley method \cite{bal86}, and a typical solution
(conformal valley) \cite{yung88}
gives a monotonous function of the conformal parameter, which turns
to zero at $R\rightarrow 0$. In the traditional language, the
valley approach corresponds to the summation of all so called
"soft-soft" corrections arising from the particle interaction
in the final state. Main problem is in the evaluation of
 "hard-hard" corrections \cite{MU91}, which
come from particle interaction in the initial state.
These corrections are likely to decrease the cross section,
and in physical terms
must take into account an (exponentially small) overlap
between the initial state, which involves a few hard quanta, with the
semiclassical final state \cite{banks}.
Thus, the instanton-antiinstanton action is
substituted by an effective "holy grail" function, which
determines the leading exponential factor for the
semiclassical production at high energies, and
which received a lot of attention in recent years.
Unitarity arguments \cite{zakh91,magg91} suggest that
the decrease of the action will stop at values of order
$S(\xi)\simeq 0.5$. In a recent preprint \cite{DP93}
Diakonov and Petrov argue that $S(\xi)$ indeed decreases up
to the value
$1/2$ at a certain energy of order of the sphaleron mass, and
then starts to increase, so that the semiclassical production
cross section is resonance-like. The question seems to us
to be not settled finally. In this study, we have taken the
value $S=1/2$ as a reasonable guess for the residual suppression,
and assumed that the behavior of the "true" function $S(\xi)$
for $S(\xi)>1/2$ is close to that given by the conformal valley
\cite{yung88}. The latter assumption is supported by numerical
studies, e.g. in \cite{DP93}.

We shall see below that
 the leading contribution
in the strong coupling comes from the following regions of integration:
\begin{eqnarray}
      z^2 &\sim & 1/(Q^2\alpha_s)
 \nonumber\\
     (z-R-T)^2+\rho_1^2 &\sim &  T^2 +\rho_2^2\sim z^2/\alpha_s
\nonumber\\
    (z-R-T)^2 &\sim & T^2 \sim  R^2 \sim \rho_1^2\sim
\rho_1^2 \sim  z^2/\alpha_s
\label{xap}
\end{eqnarray}
and additionally
$\rho^2/R^2 \sim 1-x$
when $x$ is close to $1$.
Note that these regions of integration correspond to imaginary part of
the  $I\bar{I}$ contribution so effectively the $z_i^2$ are negative.

Since $z^2$ is small we can use the lightcone expansion (see e.g.
\cite{lk}) for the quark
propagator in the $I\bar I$ background.Using the explicit expressions
for the propagators from \cite{brown} we find:
\begin{eqnarray}
\lefteqn{\bar \kappa_0(x) \sigma_\mu
 \langle x|\nabla_1^{-2} \bar
 \nabla_1\partial \bar\nabla_2 \nabla_2^{-2}
|0\rangle \sigma_\nu \kappa_0(0) =}
\\
&&\mbox{} =
-\frac{1}{2\pi^4} \int_0^1 d\gamma \frac{(\rho_1\rho_2)^{3/2}}
{[(z-R-T)^2+\rho_1^2]^2[T^2+\rho_2^2]^2}
\frac{1}{\sqrt{R^2}} \mbox{Tr}
\left\{
\frac{\bar\sigma_\nu z \bar\sigma_\mu}
     {z^4}\left[(z-R-T)
\right.\right.
 \nonumber\\
 &&\mbox{} +
\left.\left.
\rho_1^2\frac{(\gamma z-R-T)}{(\gamma z-R-T)^2}\right]
\bar R \frac{1}{\sqrt{1+\rho^2_1/(\gamma z-R-T)^2}}
 \frac{\partial}{\partial \gamma}
 \frac{1}{\sqrt{1+\rho^2_2/(\gamma z-T)^2}}
 \left[T-\rho_2^2\frac{(\gamma z-T)}{(\gamma z-T)^2}\right]
\right\} +\ldots
\nonumber
\label{prop1}
\end{eqnarray}
Omitted terms have turned out to be of order $O(\alpha_s)$.

Let us at first consider the toy example of the  Eq.\ref{formula1}
without extra integration over $\gamma$ which brings only technical
complexities:
\begin{equation}
 \int dz\frac{e^{iqz}}{z^{2n}} \int dT
\int\frac{d\rho_1^2}{\rho_1^2}(\rho_1^2)^{\mu_1}
\int\frac{d\rho_2^2}{\rho_2^2}(\rho_2^2)^{\mu_2}
\int dR e^{ipR-\frac{4\pi}{\alpha_s}(1-\frac{6}{\xi^2}) }
\frac{\Gamma(m_1)\Gamma(n)}{[(z-R-T)^2 +\rho_1^2]^{m_1} }
 \frac{\Gamma(m_2)}{[T^2 +\rho_2^2]^{m_2} }
\end{equation}

This integral diverges at $\rho\rightarrow \infty$. However,the
divergent part corresponding to instanton with size
$\rho\sim\Lambda_{QCD}$  contributes only to parton densities and not
to the coefficients in front of them. Moreover, these divergent parts
possess no imaginary part so we shall imply them subtracted in what
follows.(Strictly speaking we need $\mu-m$ subtractions of the type
$((x-R-T)^2+\rho^2)^{-2} - \rho^{-4}+2(x-R-T)^2\rho^{-6}+...$).
After performing the integrations one obtains
\begin{equation}
\pi^5\int dz\frac{\Gamma (n)\Gamma (-l)}{(z^2)^{n-l}}\int_0^1 du
\frac{u^{\mu_1+\mu_2}\bar{u}^{l-1}}{2^{m_1+m_2-1}
(1+u)^{\mu_1+\mu_2-1}}(\frac{6\pi\bar{u}^2}{\alpha_s(1+u)^2})^{\mu_1+\mu_2}
e^{ipxu-\frac{4\pi}{\alpha_s}(1-\frac{3\bar{u}^2}{2(1+u)^2})}
\label{coord}
\end{equation}
where we use the notation $l=\mu_1+\mu_2-m_1-m_2+4$ and $\bar u$ = $1-u$.
After continuation to Minkowski space the imaginary part of this
integral take the form:
\begin{equation}
z\pi^8\frac{B(n,-l)}{(Q^2)^{2-n+l}}\bar{x}^{n-2}
\frac{x^{\mu_1+\mu_2-2n+2l+2}(1+x)^{n-l-\mu_1-\mu_2}}{2^{m_1+m_2+2n-2l-3}}
(\frac{24\pi}{\alpha_s\xi^2})^{\mu_1+\mu_2-n+l}
e^{-\frac{4\pi}{\alpha_s}(1-\frac{6}{\xi^2})}
\label{mom}
\end{equation}
where $\bar x=1-x$ ,$B(n,l)=\frac{\Gamma(n)\Gamma(-l)}{\Gamma(n-l)}$
and $\xi$ is now $2(1+x)/\bar{x}$. It is easy to see now that the
 characteristic distances correspond to Eq.\ref{xap}.

One may also account for for the $\rho$ dependence in the
argument of $\alpha_s$. Careful analysis \cite{bal93b} shows that
 $\frac{4\pi}{\alpha_s}$ should be changed to
$\frac{4\pi}{\alpha_s}+2b$ ( where $b={11\over 3} N_c - {2\over 3}n_f$) and the
argument of $\alpha_s$ obeys the equation
\begin{equation}
\rho_{\ast}=
{4\pi\over \alpha_s(\rho_{\ast})}\frac{12(\xi_{\ast}-2)}{Q\xi_{\ast}^2}
\label{saddle}
\end{equation}
A numerical solution of this equation
for the particular expression of the action $S(\xi)$ corresponding
to the conformal instanton-antiinstanton
valley is shown in Fig.2.
\begin{figure}
    \begin{center}
          \begin{picture}(100,150)
          \end{picture}
    \end{center}
    \caption[xxx]{ \small
                  The non-perturbative scale in deep inelastic
                   scattering  (instanton size $\rho_\ast^{-1}$),
                  corresponding to the solution of
                  equation Eq.\ref{saddle} as a
                  function of $Q$ and for $S(\xi_\ast)\sim 0.5-0.6$
                  ($\xi_\ast\sim 3-4$).
 }
   \label{pic.b}
\end{figure}
Note that the difference between the hard scale $Q^2$ and the
effective scale for nonperturbative effects $\rho^{-2}_\ast$ is
numerically very large.  This is a new situation compared to
 calculations of instanton-induced
 contributions to two-point correlation functions, see
e.g. \cite{andrei,NSVZ80,DS80}, where the
size of the instanton is of order of the large virtuality.
The effect is
 that the instanton-induced contributions to deep
inelastic scattering may turn out to be non-negligible at the
values
$Q^2 \sim 1000 GeV^2$, which are conventionally
considered as a safe domain for perturbative QCD.

Now, in order to find the $\gamma_{\ast}g$ amplitude Eq.\ref{formula1}
we should take the real answer Eq.\ref{prop1} instead of our toy example.
After a considerable algebra (for details see \cite{bal93b})
we obtain the
following answer for the $\bar I I$ contribution
to the structure function of a real gluon:
\begin{eqnarray}
F_1^{(G)}(x,Q^2) &=&\sum_q e^2_q
\frac{1}{9\bar x^2}
\frac{ d^2\pi^{9/2}}{bS(\xi_\ast)[bS(\xi_\ast)-1]}
\left(\frac{16}{\xi_\ast^3}\right)^{n_f-3}
\nonumber\\
&&\mbox{}\times
\left(\frac{2\pi}{\alpha_s(\rho_\ast)}\right)^{19/2}
\!\!\exp\left[ -
\left(
\frac{4\pi}{\alpha_s(\rho_\ast)} +2b\right)
S(\xi_\ast)\right].
\label{answer}
\end{eqnarray}
To our accuracy, we find that the instanton-
induced contributions obey the Callan-Gross relation
$F_2(x,Q^2)= 2x F_1(x,Q^2)$.

The expression in Eq.\ref{answer} presents our main result.
It gives  the exponential correction to the coefficient
function in front of the gluon distribution of the leading twist
in Eq.\ref{cpert}.
The exponential factor is exact to the accuracy of Eq.\ref{action}.
The preexponential factor
is calculated to leading accuracy
in the strong coupling and up to corrections of order $O(1-x)$.
 The corresponding
contribution to the structure function of the nucleon is obtained
in a usual way, making a convolution of (\ref{answer}) with a
distribution of gluons in the proton at the scale $\rho_\ast^2$.

%

The instanton-antiinstanton contribution to the structure function
of a quark contains a similar contribution shown in Fig.1b.
The answer reads
\begin{eqnarray}
F_1^{(q)}(x,Q^2) &=&
\left[\sum_{q'\neq q} e^2_{q'} +\frac{1}{2}e^2_q\right]
 \frac{128}{81\bar x^3}
\frac{d^2\pi^{9/2} }{bS(\xi_\ast)[bS(\xi_\ast)-1]}
\left(\frac{16}{\xi_\ast^3}\right)^{n_f-3}
\nonumber\\
&& \mbox{}\times
\left(\frac{2\pi}{\alpha_s(\rho_\ast)}\right)^{15/2}
\exp\left[ -
\left(
\frac{4\pi}{\alpha_s(\rho_\ast)} +2b\right)
S(\xi_\ast)\right]
\label{qanswer}
\end{eqnarray}
However, in this case additional contributions exist of the
type shown in Fig.1c. They are finite (the integral over
instanton size is cut off at $\rho^2 \sim x^2/\alpha_s$),
 but the relevant instanton-antiinstanton separation
 $R$ is small, of order $\rho$.
This probably means that the structure of nonperturbative
contributions to quark distributions is more complicated.
This question is under study.
The answer given in Eq.\ref{qanswer} presents the
contribution of the particular saddle point in Eq.\ref{saddle}.
\vglue 0.4cm
{\elevenbf\noindent 3. Value of cross section and structure of final
state for instanton-induced particle production}
\vglue 0.4cm
The instanton-induced contribution to the structure function of a
gluon in Eq.\ref{answer} is shown as a function of Bjorken x for
different values of $Q\sim 10-100 GeV$ in Fig.3.
The contribution of
the box graph  is shown by dots for comparison.
The low boundary for possible values of $Q$ is determined by
the condition that the effective instanton size is not too
large. At $Q=10 \,GeV$ we find $\rho_\ast \simeq 1\, GeV^{-1}$, cf.
Fig.2. This value is sufficiently small, so that instantons
are not distorted too strongly by large-scale vacuum fluctuations.
Another limitation is that the valley approach to the
calculation of the "holy grail" function $S(\xi)$ is likely to be
justified at $S(\xi) \ge 1/2 $, which
translates to the condition that $x>0.3-0.35$.
\begin{figure}[t]
    \begin{center}
\begin{picture}(150,300)
\end{picture}
    \end{center}
\caption[xxx]{  \small
                  Nonperturbative contribution to the structure function
                  $F_1(x,Q^2)$ of a real gluon (\ref{answer})
                  as a function of $x$ for
                   different values of $Q$ (solid curves).
                   The leading perturbative contribution
                   is shown for comparison by dots. The dashed curves
                   show lines with the constant effective value of
                   the action on the $\bar I I$ configuration.
 }
   \label{pic.c}
\end{figure} \nopagebreak
Numerical results are strongly sensitive to the particular value
of the QCD scale parameter.
 We use the two-loop expression for the coupling
with three active flavors,
and the value $\Lambda_{\overline MS}^{(3)} = 365 MeV$ which
corresponds to the coupling at the scale of $\tau$-lepton mass
$\alpha_s (m_\tau)= 0.33$ \cite{ALEPH}.
Since the dependence on the coupling is exponential, the 20\% increase
of $\alpha_s(\rho_\ast)$ induces the increase of the cross section
by almost an order of magnitude! Together with uncertainties in the function
$S(\xi)$ and in the preexponential factor, this indicates that the
particular curves given in Fig.3 should not be taken too seriously,
and rather give a target for further theoretical (and experimental?)
studies to shoot at.

To summarize, we have found that
 instantons produce a well-defined and calculable
contribution to the cross section of deep inelastic scattering
 for sufficiently large values of x and large $Q^2\sim 100 - 1000
GeV^2$,
which turns out, however, to be rather small ---
of order $10^{-2}-10^{-5}$ compared to the
perturbative cross section.
This  means that the
accuracy of standard perturbative analysis is sufficiently
high, and that there is not much hope to observe
the instanton-induced contributions to the total deep inelastic
cross section experimentally.
However, instantons are likely to produce events with
a very specific structure of the final state, and such
peculiarities may be subject to experimental search.
The dominating Feynman diagrams in our calculation correspond to
 $2\pi/\alpha(\rho_\ast)\sim 15 $ gluons and $2n_f-1 =5$ quarks
in the final state with the low energy of order $\rho_\ast^{-1}
\sim 1\,GeV$. They  are produced in the spherically symmetric way
 in the c.m. frame of the partons colliding
through the instanton.( The transverse momentum of the quark
coming to the instanton is $k_{\perp}^2\sim Q^2\alpha_s \sim $ few Gev
 and so is the transverse momentum of the current quark jet).
It is not likely that quarks and gluons emitted from the instanton
can be resolved as minijets (they have  $k_{\perp}\sim Q\alpha_s \sim
$ 1 Gev), and we rather expect a spherically symmetric production of
final state hadrons in this frame. The effect is likely to be
resonance-like, that is present in  a narrow interval of values
of Bjorken x of order 0.25--0.35 (in the parton-parton collision).
In any case, finding of an instanton-induced particle production
at high energies is a challenging problem, and further theoretical
efforts are needed to put it as a practical proposal to
experimentalists.

\vglue 0.5cm
{\elevenbf \noindent 5. Acknowledgements \hfil}
\vglue 0.4cm
I would like to thank V.M.Braun for a very rewarding collaboration.
It is a pleasure to
acknowledge also useful discussions with
J.C.Collins,E.M.Levin,A.H.Mueller, and
M.I. Strikman.This work was supported by the US Department of Energy
under the grant DE-FG02-90ER-40577.
\vglue 0.5cm
{\elevenbf\noindent 6. References \hfil}
\vglue 0.4cm

\vglue 0.5cm
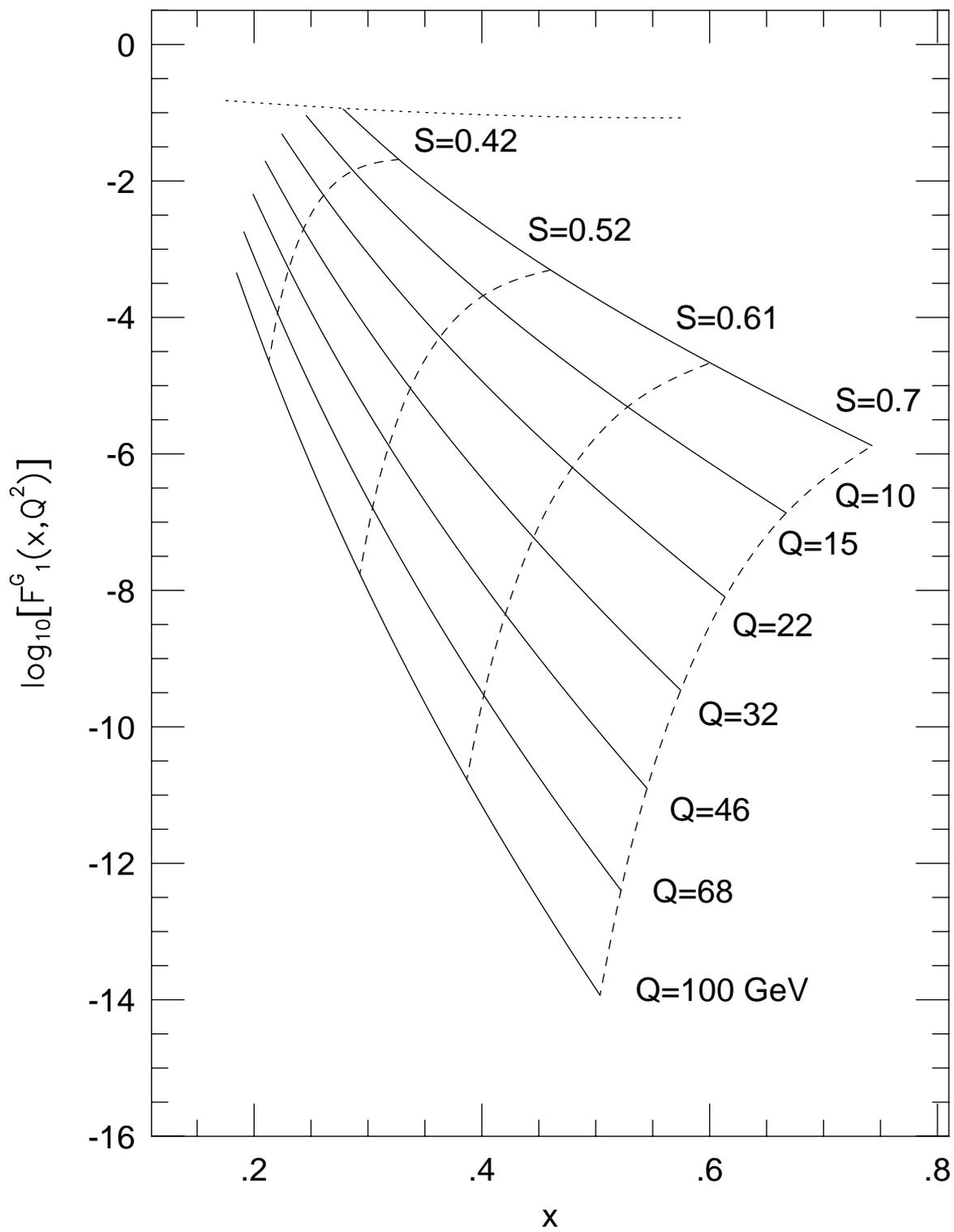
\end{document}